\documentclass[floatfix,rmp,twocolumn,twoside]{revtex4}

\usepackage{amssymb}
\usepackage{amsmath}
\usepackage[english]{babel}
\usepackage{graphicx}
\usepackage{multirow}

\makeatletter
\g@addto@macro\@verbatim\small
\makeatother

\bibpunct{[}{]}{,}{n}{}{,}

\begin{document}

\title{A Model-Driven Parser Generator, \\
from Abstract Syntax Trees to Abstract Syntax Graphs}
\author{Luis~Quesada, Fernando~Berzal, and Juan Carlos~Cubero\\
  Department of Computer Science and Artificial Intelligence, CITIC, University of Granada, \\
  Granada 18071, Spain \\
  \textit{lquesada@decsai.ugr.es, fberzal@decsai.ugr.es, jc.cubero@decsai.ugr.es}
  }

\begin{abstract}
Model-based parser generators decouple language specification from language processing. The model-driven approach avoids the limitations that conventional parser generators impose on the language designer. Conventional tools require the designed language grammar to conform to the specific kind of grammar supported by the particular parser generator (being  LL and LR parser generators the most common). Model-driven parser generators, like ModelCC, do not require a grammar specification, since that grammar can be automatically derived from the language model and, if needed, adapted to conform to the requirements of the given kind of parser, all of this without interfering with the conceptual design of the language and its associated applications. Moreover, model-driven tools such as ModelCC are able to automatically resolve references between language elements, hence producing abstract syntax graphs instead of abstract syntax trees as the result of the parsing process. Such graphs are not confined to directed acyclic graphs and they can contain cycles, since ModelCC supports anaphoric, cataphoric, and recursive references.
\end{abstract}

\maketitle

\section{Introduction} \label{sec:introduction}
A formal language represents a set of strings \cite{Jurafsky2009}.
Formal languages consist of an alphabet, which describes the basic symbol or character set of the language, and a grammar, which describes how to write valid sentences of the language \cite{Ginsburg1975,Harrison1978}.
In Computer Science, formal languages are used, among other things, for the precise definition of data formats and the syntax of programming languages.

Most existing language specification techniques \cite{Aho2006} require the language designer to provide a textual specification of the language grammar. The proper specification of such a grammar is a nontrivial process that depends on the lexical and syntax analysis techniques to be used, since each kind of technique requires the grammar to comply with a specific set of constraints. Each analysis technique is characterized by its expression power and this expression power determines whether a given analysis technique is suitable for a particular language. The most significant constraints on formal language specification originate from the need to consider context-sensitivity, the need to perform an efficient analysis, and some techniques' inability to resolve conflicts caused by grammar ambiguities.

As an alternative approach, model-based language specification techniques \cite{Kleppe2007} decouple language design from language processing and automatically generate the corresponding language grammar, thus making the language design process less arduous.

While, in general, the result of the parsing process is an abstract syntax tree that corresponds to a valid parsing of the input text according to the language concrete syntax, nothing prevents the model-based language designer from modeling non-tree structures.

Typically, syntax analysis defers some analysis tasks to later stages in the language processing pipeline, such as reference resolution and other semantic checks. However, a model-driven parser generator can be employed to automate some parts of this process.

ModelCC \cite{Quesada2011c} is a model-based parser generator that includes support for dealing with references between language elements, thus incorporating the reference resolution that is traditionally hand-crafted with the help of a symbol table into the parsing process.

In this paper, we explain how ModelCC \cite{Quesada2011c} is able to resolve references and obtain abstract syntax graphs as the result of the parsing process, rather than the traditional abstract syntax trees obtained from conventional parser generators.

Section \ref{sec:background} introduces model-based language specification.
Section \ref{sec:graphicallanguagesupport} explains the reference resolution support in the ModelCC model-based parser generator.
Section \ref{sec:exampleofagraphicallanguagemodel} includes a case study that illustrates abstract syntax graph parsing.
Finally, section \ref{sec:conclusionsandfuturework} presents our conclusions and future work.

\section{Background} \label{sec:background}

In its most general sense, a model is anything used in any way to represent something else. In such sense, a grammar is a model of the language it defines.
In Software Engineering, data models are also common. Data models explicitly determine the structure of data. Roughly speaking, they describe the elements they represent and the relationships existing among them.
From a formal point of view, it should be noted that data models and grammar-based language specifications are not equivalent, even though both of them can be used to represent data structures. A data model can express relationships a grammar-based language specification cannot.
A data model does not need to comply with the constraints a grammar-based language specification has to comply with. Typically, describing a data model is generally easier than describing the corresponding grammar-based language specification.

In practice, when we want to build a complex data structure from the contents of a file, the implementation of the mandatory language processor needed to parse the file requires the software engineer to build a grammar-based language specification for the data as represented in the file and also to implement the conversion from the parse tree returned by the parser to the desired data structure, which is an instance of the data model that describes the data in the file.

Whenever the language specification has to be modified, the language designer has to manually propagate changes throughout the entire language processor tool chain, from the specification of the grammar defining the formal language (and its adaptation to specific parsing tools) to the corresponding data model. These updates are time-consuming, tedious, and error-prone. As these changes are labor-intensive, the traditional language processing approach hampers the maintainability and evolution of the language used to represent the data \cite{Kats2010}.

Moreover, it is not uncommon for different applications to use the same language. For example, the compiler, different code generators, and other tools such as IDE editor or debugger, typically need to grapple with the full syntax of a programming language. Unfortunately, their maintenance typically requires keeping several copies of the same language specification in sync.

The idea behind model-based language specification is that, starting from a single abstract syntax model (ASM) that represents the core concepts in a language, language designers can develop one or several concrete syntax models (CSMs). These CSMs can suit the specific needs of the desired textual or graphical representation. The ASM-CSM mapping can be performed, for instance, by annotating the abstract syntax model with the constraints needed to transform the elements in the abstract syntax into their concrete representation.

This way, the ASM representing the language can be modified as needed without having to worry about the language processor and the peculiarities of the chosen parsing technique, since the corresponding language processor will be automatically updated.

Finally, as the ASM is not bound to a particular parsing technique, evaluating alternative and/or complementary parsing techniques is possible without having to propagate their constraints into the language model. Therefore, by using an annotated ASM, model-based language specification completely decouples language specification from language processing, which can be performed using whichever parsing techniques are suitable for the formal language implicitly defined by the abstract model and its concrete mapping.

\begin{figure}[h!]
\centering
\includegraphics[scale=0.225]{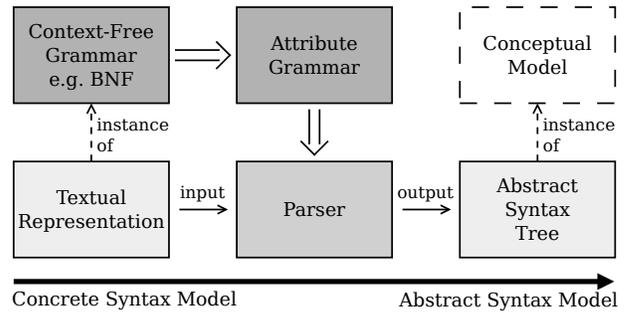}
\caption{Traditional language processing.} \label{fig:traditional}
\end{figure}

\begin{figure}[h!]
\centering
\includegraphics[scale=0.225]{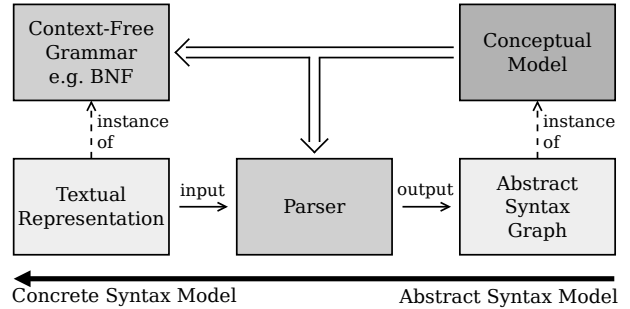}
\caption{Model-based language processing.} \label{fig:ModelCC}
\end{figure}

A diagram summarizing the traditional language design process is shown in Figure \ref{fig:traditional}, whereas the corresponding diagram for the model-based approach is shown in Figure \ref{fig:ModelCC}.

\begin{table*}[tb]
\begin{center}

\setlength{\tabcolsep}{5pt}
\begin{tabular}{ l  l  l } \hline

Constraints on... & Annotation & Function \\ \hline

\multirow{2}{*}{Patterns}
& @Pattern & Pattern matching definition of basic language elements. \\
& @Value & Field where the recognized input element will be stored. \\ \hline

\multirow{3}{*}{Delimiters}
& @Prefix & Element prefix(es). \\
& @Suffix & Element suffix(es). \\
& @Separator & Element separator(s). \\ \hline

\multirow{3}{*}{Cardinality}
& @Optional & Optional elements.\\
& @Minimum & Minimum element multiplicity.\\
& @Maximum & Maximum element multiplicity.\\ \hline

\multirow{3}{50pt}{Evaluation order}
& @Associativity & Element associativity (e.g. left-to-right). \\
& @Composition & Eager or lazy composition for nested composites. \\
& @Priority & Element precedence. \\ \hline
\end{tabular}
\end{center}
\caption{Summary of the basic metadata annotations supported by ModelCC.} \label{fig:tablesummary}
\end{table*}

It should be noted that ASMs may represent non-tree structures. Hence the use of the `abstract syntax graph' term in Figure \ref{fig:ModelCC}.

ModelCC \cite{Quesada2011c} is a parser generator that supports a model-based approach to the design of language processing systems.
Its starting ASM is created by defining classes that represent language elements and establishing relationships among those elements. Once the ASM is established, constraints can be imposed over language elements and their relationships as annotations in order to produce the desired ASM-CSM mapping.

The ASM is built on top of basic language elements, which can be viewed as the tokens in the model-driven specification of a language. ModelCC provides the necessary mechanisms to combine those basic elements into more complex language constructs, which correspond to the use of concatenation, selection, and repetition in the syntax-driven specification of languages.

In ModelCC, the constraints imposed over ASMs to define a particular ASM-CSM mapping are declared as metadata annotations on the model itself. Now supported by all the major programming platforms, metadata annotations are often used in reflective programming and code generation \cite{Fowler2002}. Table \ref{fig:tablesummary} summarizes the set of constraints supported by ModelCC for establishing ASM-CSM mappings between ASMs and their concrete representation in textual CSMs.

When the ASM represents a tree-like structure, a model-based parser generator is equivalent to a traditional grammar-based parser generator in terms of expression power. When the ASM represents non-tree structures, reference resolution techniques can be employed to make model-based parser generators more powerful than grammar-based ones, as we will see in the next Section.

\section{Reference Resolution Support in ModelCC} \label{sec:graphicallanguagesupport}

Reference resolution consists of finding the object a reference refers to and, in the case of ModelCC, automatically linking the reference to the corresponding object instantiation. This resolution process is what leads to abstract syntax graphs instead of trees in model-driven language processing.

In ModelCC, an object reference is embodied by a subset of the elements in its full object definition. This subset of elements acts as an identifier (or key in database terms) that, when found in the input text, can be recognized as a reference to the corresponding object in the model and linked to its instantiation in the ASM.

References in ModelCC can be anaphoric, when they are preceded by the corresponding object definition, but also cataphoric, when the references precede the definition, and even recursive, when they appear within the definition they refer to.

Subsection \ref{sec:id} introduces the \emph{@ID} metadata annotation, which allows the specification of identifiers for language elements.
Subsection \ref{sec:ref} presents the \emph{@Reference} annotation, which allows the specification of references to other language elements.

\subsection{The @ID Annotation} \label{sec:id}
ModelCC uses an \emph{@ID} metadata annotation to support reference specification. This annotation is applied to a subset of the members of a language element model. This subset determines the syntax of references to particular instances of such elements in the concrete syntax of the corresponding language. That is, any appearance of the same set of values will be interpreted as a reference to the same instance of the referred language element.

The use of references is resolved in our implementation of ModelCC by the introduction of grammar productions that characterize such references and semantic actions that map them to the corresponding language elements.

\begin{figure}[t]
\centering
\includegraphics[scale=0.97]{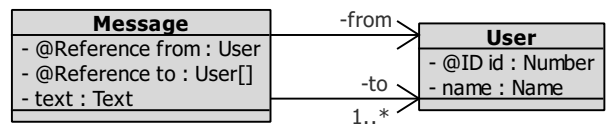}
\caption{ModelCC specification of \emph{Messages}, their senders, and their receivers.} \label{fig:graphuserref}
\end{figure}

In Figure \ref{fig:graphuserref}, the \emph{@ID} annotation is employed to identify users by a single number.

It should be noted that the \emph{@ID} annotation is incompatible with the \emph{@Optional} ModelCC annotation, as null language element identifiers are not allowed, for the same reasons that attributes in a primary key are not nullable in a relational database.

However, the \emph{@ID} annotation can be used together with other ModelCC annotations, such as \emph{@FreeOrder}, which allows the members of a language element to be shuffled in their textual representation, and \emph{@Prefix} and \emph{@Suffix}, which add syntactic sugar to the incarnation of the abstract syntax model as a concrete textual language.

The inadvertent definition of two entities of the same class with the same identifier results in a runtime warning produced by ModelCC when parsing its input.

\subsection{The @Reference annotation} \label{sec:ref}

ModelCC resorts to the \emph{@Reference} metadata annotation to complete its support for reference resolution. The \emph{@Reference} annotation applies to individual members of any language element, provided that the referenced types contain at least one \emph{@ID}-annotated member in their language model.

Whenever a language element member is annotated with \emph{@Reference}, the corresponding grammar productions are modified so that they refer to the symbol corresponding to the element reference specification rather than the symbol that corresponds to its full specification. These productions are then associated to a semantic action that resolves the references at the end of the parsing process, in order to support cataphoric and recursive references, apart from the anaphoric references that could be resolved on the fly during the parsing process.

In Figure \ref{fig:graphuserref}, the textual syntax of messages includes numbers that, as identifiers, refer to particular users. ModelCC will parse such identifiers, recognize the references, resolve them, and return the correct object graph.

\section{A Working Case Study} \label{sec:exampleofagraphicallanguagemodel}

In this section, we present an example language that allows the specification and rendering of complex 3D objects using the reference resolution capabilities of ModelCC.

First, we will outline the features we wish to include in our 3D object specification language.
Then, we will provide the full language specification for ModelCC by defining an abstract syntax model, which will be annotated to specify the desired ASM-CSM mapping.
Lastly, we will see some examples of input and output pairs for our 3D object specification language.

\subsection{Language Description}

Our 3D object specification language is designed to support the following features:

\begin{itemize}
\item A special section, denoted by the ``scene'' keyword, delimits the statements that will be used for rendering the scene.
\item The definition of custom objects, which are identified by an object name. As references can be lazily resolved, recursion is allowed.
\item Scoped statements, delimited by ``\{'' and ``\}'', that allow the specification of lists of statements that will run sequentially in a new OpenGL scope (that is, issuing a ``glPushMatrix'' before executing the statements and ``glPopMatrix'' after executing the statements).
\item Composite statements, delimited by ``['' and ``]'', that allow the specification of lists of statements that will run sequentially, but without creating a new OpenGL scope.
\item Repeated statements that allow the repetition of a statement, a group of statements, or a block of statements, a specific number of times.
\item Object statements, which draws either basic objects (e.g. a cube) or user-defined objects. Draw statements allow the specification of a numeric parameter. The ``next'' keyword is replaced in runtime by the current parameter decreased by one, and draw statements will not run when the parameter is 0.
\item State-machine OpenGL-like scale transformation statements, which support the specification of a combination of x, y, and z values in any order, or a single scaling factor that will be applied to the three axes.
\item State-machine OpenGL-like rotate transformation statements, which support the specification of the angle and a combination of x, y, and z axis values in any order.
\item State-machine OpenGL-like translate transformation statements, that support the specification of a combination of x, y, and z values in any order.
\item State-machine color transformation statements, which support the specification of a combination of red, green, blue, and alpha values in any order, and allow either absolute (by default) or relative color adjustments.
\end{itemize}

\begin{figure*}[p!]
\centering
\includegraphics[scale=0.97]{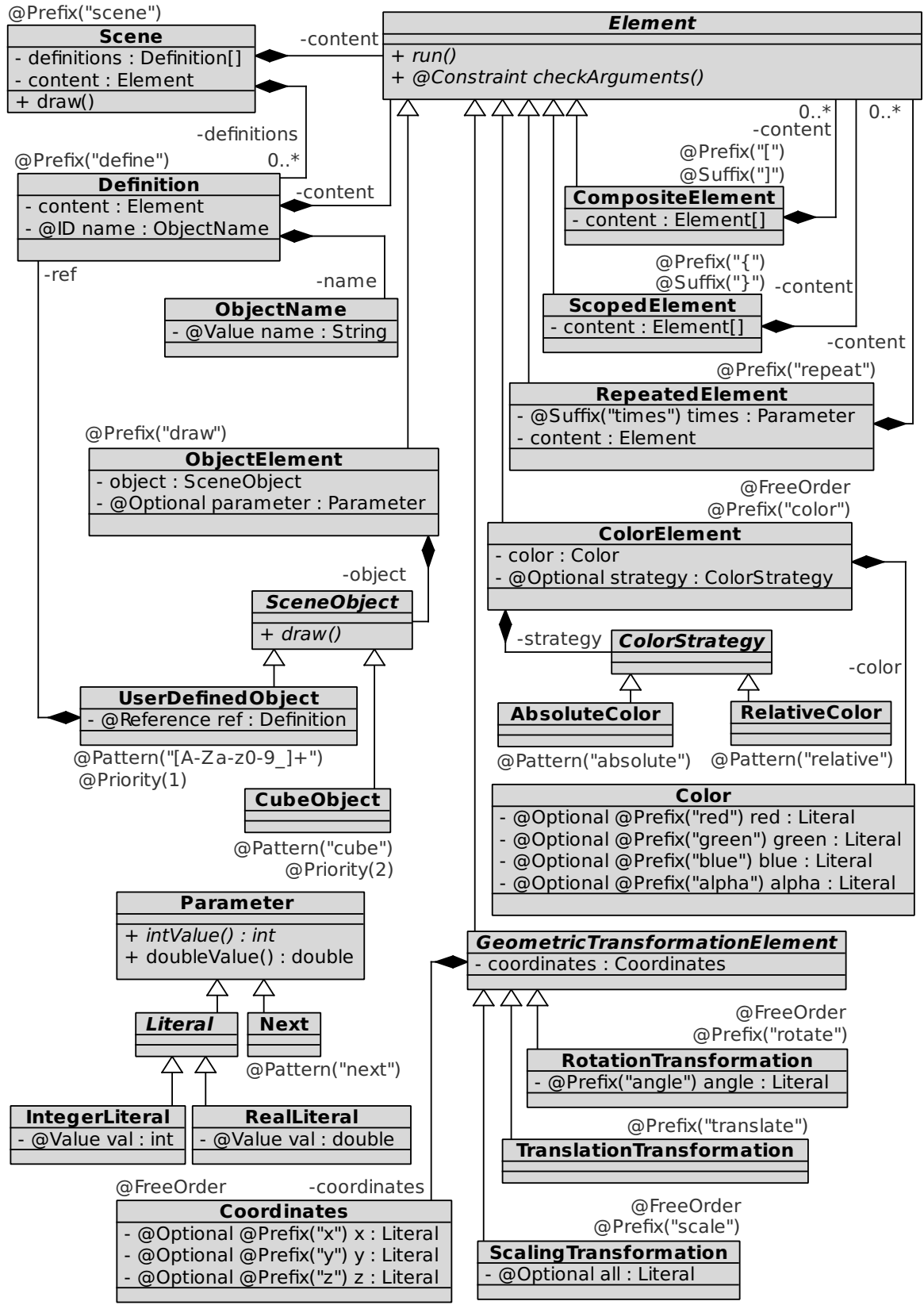}
\caption{ModelCC definition of a 3D object specification language. ModelCC reference resolution support is used to allow the specification of complex 3D objects in the \emph{Definition} class.} \label{fig:graphmodelcc}
\end{figure*}

\begin{figure}[p!]
\begin{verbatim}
define snail [
  draw cube
  {
    scale 0.3
    color blue 1
    repeat 6 times [
      draw cube
      translate y 1
      rotate z 1 angle -5
      color relative alpha -0.06
    ]
  }
  translate x 0.8
  rotate z 1 angle 10
  scale 0.98
  color relative
     red -0.05 green +0.05 alpha -0.008
  draw snail next
]
scene [
  color red 1
  draw snail 400
]
\end{verbatim}
\caption{Snail specification in our 3D object specification language.} \label{fig:examplecode1}
\end{figure}

\begin{figure}[p!]
\centering
\includegraphics[scale=0.21]{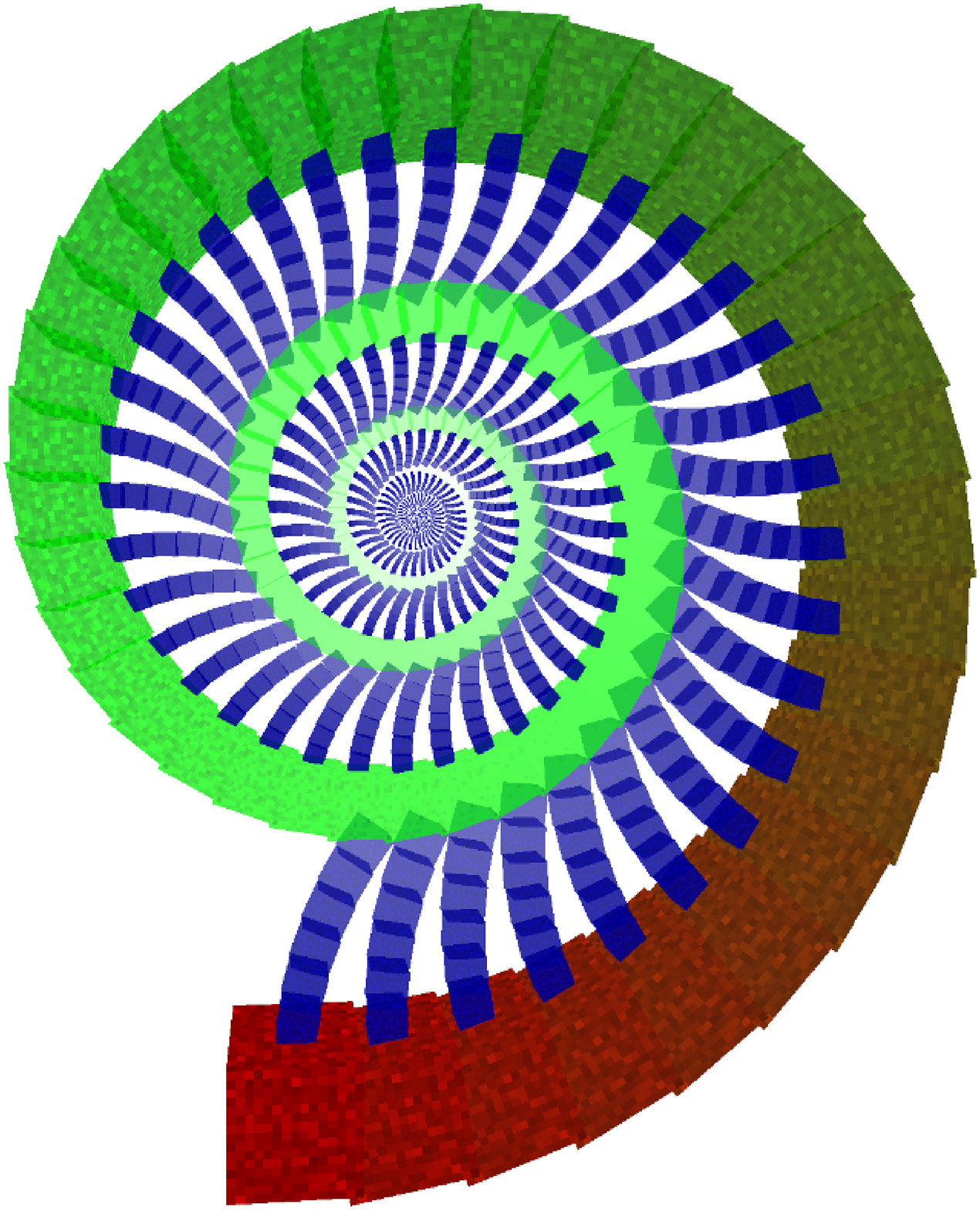} \\ [0.2cm]
\includegraphics[scale=0.21]{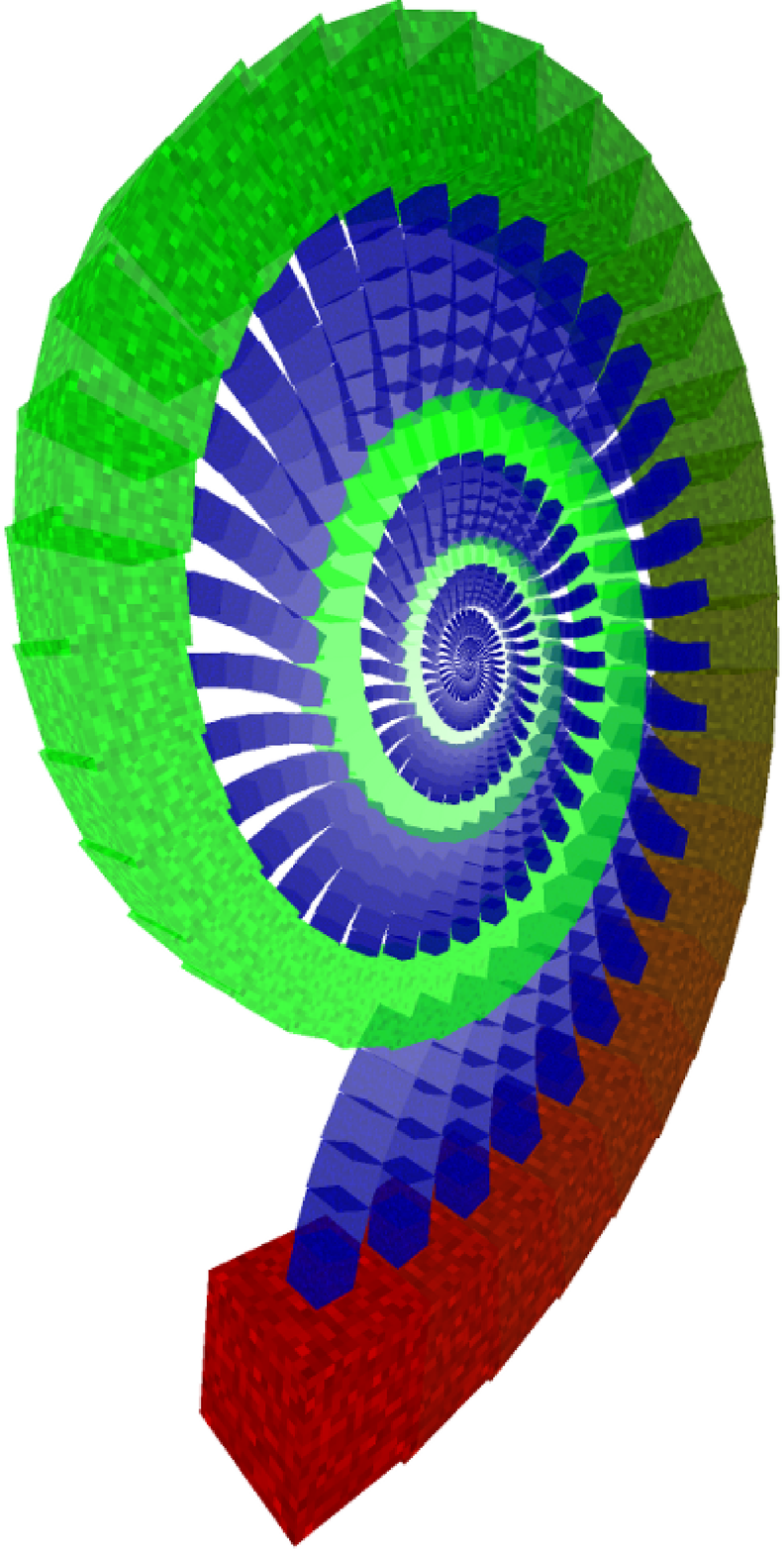}

\caption{Different views of the snail specified by the input text shown in Figure \ref{fig:examplecode1}.} \label{fig:examplefig1}
\end{figure}

\begin{figure}[p!]
\begin{verbatim}
define helix [
  {
  scale x 0.4 z 0.4
  draw cube
  }
  {
    rotate y 1 angle 45
    scale 0.4
    scale y 0.2 x 0.2 z 1.5
    repeat 10 times [
      draw cube
      color relative alpha -0.08
      translate z -1
    ]
  }
  translate y 1
  translate x -4 z -4
  rotate y 1 angle 6
  translate x 4 z 4
  draw helix next
]
scene [
  {
    rotate y 1 angle 90    color red 1
    translate x 4 z 4   draw helix 40
  } {
    rotate y 1 angle 180   color green 1
    translate x 4 z 4   draw helix 40
  } {
    rotate y 1 angle 270   color blue 1
    translate x 4 z 4   draw helix 40
  } {
    color red 0 green 0 blue 0
    translate x 4 z 4   draw helix 40
  }
]
\end{verbatim}
\caption{Quadruple helix specification in our 3D object specification language.} \label{fig:examplecode2}
\end{figure}
\begin{figure}[p!]
\centering
\includegraphics[scale=0.10]{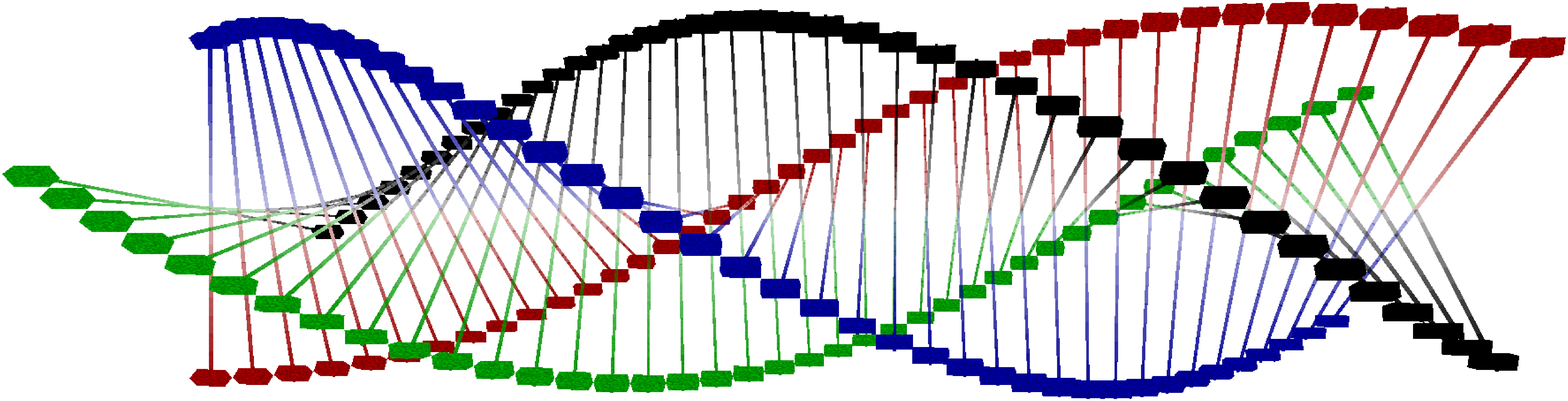} \\ [0.2cm]
\includegraphics[scale=0.30]{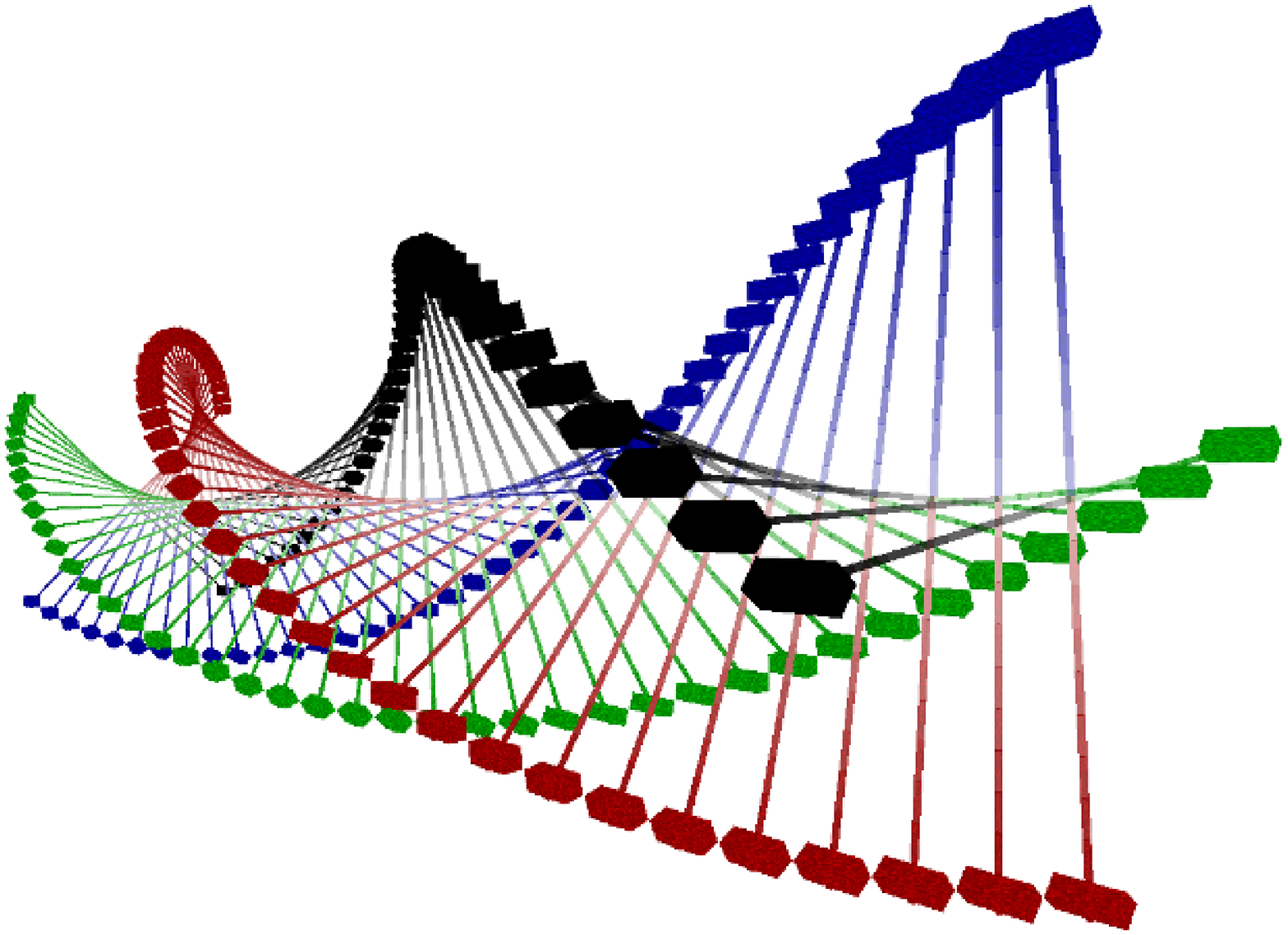}

\caption{Different views of the quadruple helix specified by the input text shown in \ref{fig:examplecode2}.} \label{fig:examplefig2}
\end{figure}
\subsection{ModelCC Implementation}

In ModelCC, the abstract syntax model is designed first and then it is mapped to a concrete syntax model by imposing constraints by means of metadata annotations on the abstract syntax model.

The resulting model can be processed by ModelCC to generate the corresponding parser.
The UML class diagram in Figure \ref{fig:graphmodelcc} presents our annotated 3D object specification language model.

The reference support extension we propose in this paper can be observed in the \emph{Definition}, \emph{ObjectName}, and \emph{DefinedObject} classes. The \emph{name} member of the \emph{Definition} class is annotated with \emph{@ID}, which means that a \emph{Definition} instance can be identified by an \emph{ObjectName}. Then, the \emph{ref} member of a \emph{DefinedObject} is annotated with \emph{@Reference}, which means that, in textual form, a \emph{DefinedObject} can refer to a \emph{Definition} by its \emph{ObjectName}. ModelCC reference resolution allows references to be resolved during the parsing process and makes the implementation of a traditional symbol table unnecessary.

It should be noted that certain constraints cannot be expressed in the abstract syntax model. However, these constraints can be expressed as custom constraints using the \emph{@Constraint} annotation. In our example, some statements corresponding to elements in our model, such as draw statements and repeat statements, will not accept real values as parameters. These custom semantic constraints are implemented in the \emph{checkArguments()} methods of the language elements classes corresponding to those statements.

ModelCC is able to automatically generate a grammar from the ASM defined by a class model and the ASM-CSM mapping defined as a set of metadata annotations on the class model. References in that grammar are automatically resolved by ModelCC so that further work is not needed.

\subsection{Examples of 3D Object Specification}

Figures \ref{fig:examplecode1} and \ref{fig:examplefig1} illustrate the specification and rendering of a 3D snail in our 3D object specification language. The \emph{snail} object is defined as a single section of the snail consisting of the shell and a blue strip, and a slightly smaller, more transparent, and more greenish snail. The scene consists of a 400-section \emph{snail} object.

Figures \ref{fig:examplecode2} and \ref{fig:examplefig2} illustrate the specification and rendering of a quadruple 3D helix in our 3D object specification language. The \emph{helix} object is defined as a single section of a helix consisting of the outer part and a strip that grows more transparent until it reaches the axis, and a slightly rotated and translated helix. The scene consists of red, green, blue, and black 40-section \emph{helix} objects.

\section{Conclusions and Future Work} \label{sec:conclusionsandfuturework}

ModelCC is a model-based parser generator that employs metadata annotations to implement ASM-CSM mappings.

We have described how ModelCC supports reference resolution and allows parsing abstract syntax graphs rather than conventional abstract syntax trees, as obtained by traditional grammar-driven parser generators.

We have demonstrated the use of ModelCC reference resolution support by designing and implementing a fully-functional abstract syntax graph parser for a 3D object specification language.

In the future, we plan to apply model-based language specification techniques to problems such as data integration. We also plan to implement metadata annotations that support more complex scoping rules for reference resolution.

\renewcommand{\baselinestretch}{0.98}
\bibliographystyle{plain}
{\small
\bibliography{doc}}
\renewcommand{\baselinestretch}{1}

\end{document}